# A Short Working Distance Multiple Crystal X-ray Spectrometer

B. Dickinson[1], G.T. Seidler[1,*], Z. W. Webb[1], J.A. Bradley[1], K.P. Nagle[1], S.M. Heald[2], R.A. Gordon[3], and I.M. Chou[4]

[1] Physics Department, University of Washington, Seattle, Washington, 98195
[2] Advanced Photon Source, Argonne National Labs, Argonne, Illinois, 60439
[3] Physics Department, Simon Fraser University, BC, V5A 1S6, Canada
[4] U.S. Geological Survey, Reston, VA 20192

For x-ray spot sizes of a few tens of microns or smaller, a mm-sized flat analyzer crystal placed ~ 1 cm from the sample will exhibit high energy resolution while subtending a collection solid angle comparable to that of a typical spherically bent crystal analyzer (SBCA) at much larger working distances. Based on this observation and a non-focusing geometry for the analyzer optic, we have constructed and tested a short working distance (SWD) multicrystal x-ray spectrometer. This prototype instrument has a maximum effective collection solid angle of 0.14 sr, comparable to that of 17 SBCA at 1 meter working distance. We find good agreement with prior work for measurements of the Mn $K\beta$ x-ray emission and resonant inelastic x-ray scattering (RIXS) for MnO and also for measurements of the x-ray absorption near-edge structure for Dy metal using $L\alpha_2$ partial-fluorescence yield detection. We discuss future applications at third- and fourth-generation light sources. For concentrated samples, the extremely large collection angle of SWD spectrometers will permit collection of high-resolution x-ray emission spectra with a single pulse of the Linac Coherent Light Source.

(*) corresponding author: seidler@phys.washington.edu

**I.** Introduction

The efficient and precise measurement of x-ray emission spectra is important to many branches of chemistry, biology, condensed matter physics, plasma physics, and astrophysics. Continuing developments in the performance of low-noise x-ray position sensitive detectors have led to a parallel series of innovations in spectrometer design.[1-7] Of most relevance for the present paper, these detectors greatly increase the efficiency of dispersing spectrometer configurations when detector pixels are illuminated by photons having a narrow range of analyzed energies. Typical examples include spectrometers based on the von Hamos geometry,[8] commonly used in laser-induced plasma research,[5] and on the Johann geometry,[9] which has many applications at synchrotron light sources.[1,4,7]

We propose an alternative, but related design for dispersive spectrometers for x-ray emission spectroscopy applications where a small source size is available, such as at synchrotron light sources or in analytic electron microscopy. For x-ray spot sizes of a few tens of microns or smaller, a mm-sized unstrained flat analyzer crystal placed ~ 1 cm from the sample provides high energy resolution while subtending a collection solid angle comparable to that for a typical spherically bent-crystal analyzer (SBCA) at much larger working distances. An appropriately arranged ensemble of such crystals at short working distance (SWD) thus forms the basis for an x-ray spectrometer having an extremely large collection solid-angle, subject to either of two conditions. First, the analyzed radiation from all the dispersing elements may superimpose usefully (*e.g.*, focus at a line or plane). This holds at longer working distances for diced Johann analyzers[4] or for crystals on a Johann surface of revolution,[10] and also has been demonstrated for SWD 'multi-stepped' optics consisting of flat or curved elements.[11-13] Alternatively, as we illustrate here, a useful spectrometer can instead be constructed when the analyzed radiation from all the dispersing elements is *completely non-overlapping* on some plane so that each pixel of a 2-dimensional position-sensitive detector (2D-PSD) is uniquely illuminated. When the energy range scattered by the analyzer crystals onto the detector matches the energy range of physical interest, the focusing and nonfocusing approaches have equal efficiency per collection solid angle. The purpose of this paper is to demonstrate that this simplest approach to multiplexing the flat-crystal analyzer is both straightforward to implement and extremely effective. The analysis of the reflection from each independent crystal is similar to that for existing x-ray spectrometers using single flat crystals and 2D-PSD's at third- and fourth-generation light sources.[2,14]

The key parameters that determine the energy resolution of a dispersive spectrometer are the angular resolution $\delta\theta_{crystal}$ of the dispersing element, and the angular sizes $\delta\theta_{source}$ and $\delta\theta_{pixel}$ of the source (or spot on the sample) and the detector pixels, respectively. It is useful to first consider a 'primitive' spectrometer having $\delta\theta_{crystal} \sim \delta\theta_{source} \sim \delta\theta_{pixel}$ and consisting of a microfocused beam, a single unstrained flat crystal analyzer, and a 2D-PSD. For the sake of discussion, we take $\delta\theta_{crystal} = 10^{-4}\,\text{rad}$, such as would be appropriate for Si reflections near backscatter for energies of 5-10 keV. The resulting spectrometer is shown in Figure 1a for a 1 μm incident spot size on the sample and a typical pixel size of 100-200 μm. This instrument's combination of extremely high energy resolution and large bandwidth is not, in practice, a beneficial coexistence. It is instead necessary to find a compromise which provides a desired analyzer energy range and effective energy resolution for a particular experiment while still maintaining a large collection solid angle.

We focus here on obtaining energy resolutions of a few tenths to a few eV and detector bandwidths of 10-50 eV for hard x-ray applications, *i.e.*, for use in measurement of *Kα* or *Kβ* fluorescence for 3*d* transition metals or for *Lα* or *Lβ* fluorescence from rare earths or actinides.  When operating relatively near to backscatter, such a spectrometer will have a source-to-analyzer distance of a few cm and an analyzer-to-detector distance of 5-20 cm.  An instrument using only one flat dispersing element is useful for some applications,[2] but it is natural to consider using multiple crystals to increase the effective solid angle available for the 2D-PSD. A few subtleties arise in such designs. While an approximate focal plane exists for diced Johann analyzers subtending a relatively small angle,[4] this is not the case for optics with the larger angular size of interest here. Further, as shown in Figure 1b, unique illumination of the 2D-PSD in the far-field does not occur when two flat crystals are placed next to each other on the usual circle for the Johann geometry. This problem is solved by an appropriate aperture (Figure 1c) that permits a significant fraction of the total solid angle of the dispersive optic to uniquely illuminate the 2D-PSD, thus forming an efficient x-ray spectrometer. One could alternatively have a gap between successive crystal elements or mask the crystals to achieve the same purpose.

The design in Fig. 1c serves as a general template for a high-efficiency instrument with good energy resolution for x-ray emission spectroscopy and related techniques. While it is tempting to describe the SWD multicrystal analyzer in Fig. 1c as a coarsely diced, miniature Johann optic, this misses an important distinction: the present optic is not used in a focusing mode with respect to the detector. The radiation scattered from the respective analyzer elements specifically must *not* overlap on the detector. This detail has an important practical benefit: an instrument based on Fig. 1c has relatively weak specifications on assembly and operation. Analyzer crystal orientation errors of ~1 degree, or sample and beam placement errors of ~1 mm, will modify the location of the various backscattered patterns on the 2D-PSD and will modestly change their respective bandpasses, but will not strongly influence final spectrometer performance. Such errors are instead accomodated by an *in situ* detector calibration procedure, as demonstrated below.

This paper proceeds as follows. We provide experimental details in section II. This includes the design of an inexpensive prototype SWD multicrystal spectrometer with a large collection solid angle, potentially comparable to that of as many as 17 typical SBCA, depending on the application. In section III.A., we demonstrate a high-throughput Mn $K\beta$ resonant inelastic x-ray scattering (RIXS) study of MnO. Good-quality Mn $K\beta$ emission spectra were acquired in 40 seconds per incident energy at a third-generation synchrotron light source, and straightforward modifications to the experiment would permit an order-of magnitude decrease in measurement time. In section III.B, we present the results of a Dy $L_3$ x-ray absorption near-edge structure (XANES) experiment using partial fluorescence yield detection with fine energy resolution at the Dy $L\alpha_2$ fluorescence line. Such conditions lead to suppression of broadening from the short core-hole lifetime. Future directions are discussed in section III.C, with an emphasis on likely applications at third- and fourth-generation light sources and the possible use of a alternative geometries for improved performance over a wider analyzer energy range or more efficient use of the solid angle of the analyzer optic. It is noteworthy that even an instrument based on our present Johann-style approach will provide x-ray emission spectra with good counting statistics for relatively concentrated samples for a single pulse at the Linac Coherent Light Source. In section IV, we conclude.

**II.** Experimental

**II. A.** SWD Spectrometer Design

A cut-away rendering of our short working distance (SWD) multicrystal x-ray spectrometer is shown in Figure 2. The body of the spectrometer consists of commercial lens tube components (Thorlabs) and an Al 6061-alloy plate, which both defines the necessary spatial filtering aperture (4.5 mm diameter) and also serves to support the sample mounting process. The aperture is on the lens-tube cylindrical axis at a distance of 1-2 mm below the nominal beam path. Beam entrance and exit holes of 3-mm diameter are diametrically opposite each other on the larger lens tube; the beam path is normal to the cylindrical axis of the lens tube. A gas fitting for the He inlet (not shown) is located on the adapting piece between the two lens tubes. The open end of the larger lens tube is covered and sealed with 50-micron thick Kapton film. A small pinhole placed in this film serves as an outlet for a slow gas flow during measurements. A commercial 2D-PSD (Pilatus 100k) is placed 15 mm behind the open end of the large-diameter lens tube. The detector pixels are 172 x 172 $\mu m^2$ on a 487 x 195 grid. A face-plate with a cylindrical extension serves to mate the 2D-PSD to the spectrometer output end while also rejecting stray x-rays at the experimental station.

A photograph of the diffractive component is shown in Figure 3a. Twenty-two Si 440 crystals with dimensions 4.0 x 4.2 mm are affixed with silicone vacuum grease to the curved surface of a plano-concave lens. The lens (Thorlabs) is made from BK7 glass and has a diameter of 25 mm and a radius of curvature of 25.7 mm. The sphere defined by its concave surface is within 1 mm of the intersection of the beam and the lens tube cylindrical axis. A sample placed at this spot will be in a perfect backscattering geometry from all 22 crystals.

A key feature of this spectrometer is its large collection solid angle, as this directly influences the final efficiency for data collection. It is useful to have a reference standard for comparison to traditional x-ray spectrometers; we note that a typical 10-cm diameter spherically-bent crystal analyzer (SBCA) at a 1 m working distance subtends $7.9 \times 10^{-3}$ sr, or about 1/1600 of $4\pi$ sr. Based on the above dimensions, an experiment which made full use of the backscattered radiation through the aperture from all 22 crystals would have an effective collection solid angle of 0.14 sr, or approximately equivalent to that of 17 SBCA.

A representative 2D-PSD exposure for our test instrument is presented in Figure 3b. The oval shape of each bright region is defined by the size and placement of the aperture, and the rectangular shadow protruding into each such region is the shadow of the Pb tape sample holder used in the Dy $L_3$ XANES measurement (described below).

**II.B.** Beamline Details

All measurements were performed at the x-ray microprobe endstation 20-IDB of the XOR/PNC sector at the Advanced Photon Source.[15] A double Si (111) monochromator was used, with theoretical energy resolutions (excluding detuning) of 0.85 and 1.04 eV at 6.5 and 8.0 keV, respectively. The focused beam size was 5 microns.

The short length of the Kirkpatrick-Baez (K-B) mirrors used in the present measurements captured only a portion of the incident beam for a final focused flux of $5\times10^{11}$/sec. The intensity of the incident beam was monitored with a small He gas ionization chamber placed between the K-B mirrors and the entrance to the spectrometer. The transmitted intensity was also monitored with a He-filled ionization chamber. The monochromator was calibrated with 0.1 eV precision at the Co *K*-edge prior to measurement, and no calibration drift was seen when repeating a Co *K*-edge scan immediately after the present studies were completed.

**II.C.** Sample preparation and mounting

The starting material for the MnO sample was synthetic manganosite[16] which was ground by hand to form particles ranging from a few to 10 microns. This powder was dispersed on a piece of Kapton tape, which was then mounted at a 35 degree angle of inclination to the beam. The sample was placed on the far side of the aperture from the beam in order that a wider bandwidth of x-rays would be Bragg scattered by the analyzer crystals through the aperture. Based on the x-ray transmission monitor, a grain with effective thickness of approximately 4 µm was illuminated by the beam. The beam location was approximately 1 mm away from the top-surface of the aperture plate.

The Dy sample (section IV.A.) was a 1.0 mm$^2$ piece of a 0.1-mm thick Dy foil (99.9% purity, Alfa-Aesar). This was mounted with silicone vacuum grease onto a small piece of Pb tape which was bent to form a 45-degree angle to the beam. This placed the sample well within the lateral extent of the aperture, and approximately 0.5 mm away from the top surface of the aperture plate. The shadow of the Pb tape sample-holder is the dark rectangular protrusion in each bright region in Fig. 3b.

**II.D.** Detector Calibration and Effective Collection Solid Angle

The spectrometer performance depends on the ability to determine the energy of photons reflected to each illuminated pixel on the 2D-PSD. Following standard procedure for IXS spectrometers, this can be achieved *in situ* by collecting 2D-PSD images while scanning the incident photon energy $E_1$ through the range of expected emission energies $E_2$ for the intended study. The elastic scattering from the sample will be the dominant radiation collected in such a measurement, and can therefore be used to independently calibrate the energy of photons incident on each pixel of the detector. After this calibration is complete, the counts from pixels in selected energy bandpasses can be grouped to define the detector bins for an x-ray emission spectrum.

Figures 4 and 5 illustrate some relevant details of the detector calibration for both the MnO and Dy studies. In panels (a)-(d) of each figure, representative data at the refection from one Si (440) analyzer crystal is shown for elastic scattering during the calibration studies. Panel (e) in each figure shows a contour plot of the resulting energy calibration, and panel (f) shows representative data for the respective partial fluorescence yield (PFY) measurements. Note that the calibrations do not fill the reflection through the aperture due to our choice to scan the monochromator over only the energy range of physical interest for the present studies. This has no consequences for the results presented here.

The total solid angle subtended by the region of the 2D-PSD that can be actively illuminated by the analyzer crystals in the present configuration is 0.11 sr, equivalent to nearly 14 SBCA. However, the more relevant value is the effective solid angle for pixels whose calibrated energy is of physical interest in a given experiment. For the MnO and Dy experiments this solid angle is 0.021 sr (2.7 SBCA) and 0.061 sr (7.8 SBCA), respectively.

The reduced solid angle for the MnO experiment is due to difficulties with the calibration process used in this study. As the main axis of the apparatus was aligned in the horizontal scattering plane, the elastic scattering from the horizontally-polarized incident beam is greatly reduced by the usual polarization factor. By failing to take elastic-line scans with fine enough energy spacing and long enough integration times, only 5 of the 16 aperture projections gave complete and consistent calibrations over the entire energy range of interest.

For the Dy experiment, all crystals provide useful reflections during calibration, but only a fraction of the total aperture solid angle calibrates to the chosen 6 eV bandwidth of interest. Some additional solid angle is lost to the shadow of the sample holder. By means of comparison to a Si (440) SBCA experiment aimed at observing lifetime-broadening suppression at the peak energy of a fluorescence line,[17] it is also fair to note the smaller effective solid angle of $3.1 \times 10^{-3}$ sr (0.4 SBCA) for a single 0.3 eV bandwidth at the peak of the Dy $L\alpha_2$ fluorescence.

For the MnO study, the total response function of detector pixels in a narrow range of $E_2$ at the peak energy of the Mn $K\beta_{1,3}$ fluorescence has a full-width half maximum of 1.6 eV. Given the theoretical monochromator response, we infer a spectrometer energy resolution of 1.35 eV. The total response function for detector pixels in a narrow range of $E_2$ at the peak energy of the Dy $L\alpha_2$ fluorescence has a full-width half maximum of 0.8-0.9 eV, consistent with the theoretical monochromator response at the same energy. The spectrometer resolution for the Dy experiment is therefore a few tenths of an eV and is limited by the size of the 2D-PSD pixels.

**III.** Results and Discussion

**III.A.** X-ray Emission and Resonant Inelastic X-ray Scattering from MnO

Resonant inelastic x-ray scattering (RIXS) and resonant x-ray emission spectroscopy (RXES) are seeing growing applications due to the increased availability of dedicated instrumentation and of associated theoretical treatments.[18] One of the more interesting classes of study involves the transition metal $K\beta$ fluorescence.[19] Due to the exchange interaction between the valence 3$d$ electrons and the semicore 3$p$ electrons responsible for the $K\beta$ fluorescence, the $K\beta$ spectrum is split into majority- and minority-spin contributions. RIXS measurements, or high-resolution PFY XANES measurements with appropriately selected detector bandpasses, can therefore demonstrate a strong sensitivity to the spin-dependence of the final states.[20]

MnO serves as a canonical test case.[3,20,21] The valence electronic configuration of the $Mn^{2+}$ ion is $3d^5$, which by Hund's rule would have the maximum contrast between majority- and minority-spin contributions. The results of our RIXS study of MnO are presented in Figures 6-8. In Fig. 6, we show the Mn $K\beta$ emission spectrum for an incident energy of 6558 eV. The energies, widths, and relative amplitudes of the sharp $K\beta_{1,3}$ peak and the broad $K\beta'$ satellite are in quantitative agreement with prior studies on $Mn^{2+}$ using more traditional spectrometers.

In Fig. 7, we show a contour plot of the RIXS intensity from MnO. The results are in good agreement with those of Hayashi, *et al.*,[7,21] except that the intensity of the resonant Raman scattering of the $K\beta_{1,3}$ line at the $K\beta'$ energy is much weaker here. This is due to a difference in sample preparation: the prior work used a thick pellet sample and consequently had a strong enhancement of the magnitude of the pre-edge resonant Raman scattering relative to the edge step due to self-absorption effects. In Fig. 8 we show the PFY XANES at the $K\beta_{1,3}$ and $K\beta'$ energies. The $1s \rightarrow 3d$ transition (Fig. 8, inset) is seen strongly in the minority spin channel ($K\beta_{1,3}$), as expected. The results are in good agreement with prior studies,[7,20,21] with any differences again due to smaller self-absorption effects in the present study.

A final comparison should be made between the present and prior work. Our measurement time (40 sec per $E_1$ point) compares well with the recent study[7] at SPRING-8, which admittedly used an earlier generation of 2D-PSD. As mentioned earlier, ~30% of the potential detector collection solid angle is used here due to difficulties with the detector calibration, and only ~25% of the available incident beam is used by the microprobe optics. Order of magnitude improvements in measurement times at the same energy resolution are therefore possible in the future. Alternatively, the 2D-PSD working distance may be increased by more than a factor of two while retaining the present collection solid angle but significantly improving energy resolution through the resulting decrease of $\delta\theta_{pixel}$.

**III.B.** Lifetime Broadening Suppression for Dy $L_3$ XANES

As first discussed by Hämäläinen, *et al.*,[17] broadening of XANES spectra by the core-hole lifetime $\Gamma_{core-hole}$ is not an inescapable consequence of all x-ray absorption measurements, but is instead present only when the detection bandwidth is at least as large as $\Gamma_{core-hole}$. Very narrow bandwidth measurements of the partial fluorescence yield (PFY) in XANES or RIXS studies are instead broadened only by the lifetime of the atomic shell whose electron fills the core hole.

In Figure 9 we present partial fluorescence yield XANES measurements for the Dy foil sample. Recall that the measured response function (including the monochromator bandwidth) is 0.8 eV in this measurement. The overall experimental conditions for this measurement are thus very close to that of prior work using an SBCA.[17] The emission energy ($E_2$) bandpass shifts upward by 0.5 eV between successive curves when progressing down the figure. The total measurement time for this complete set of spectra was ~40 minutes. The lowest $E_2$ measurements are essentially coincident with the central energy of the Dy $L\alpha_2$ fluorescence, which has an intrinsic broadening greater than 5 eV FWHM due to the core-hole and intermediate-state lifetimes.[17,22]

These spectra demonstrate the expected, strong suppression of the core-hole lifetime broadening in the XANES. Differences between the present results and prior work[17] are minor. The earlier study measured $E_2$ at the Dy $L\alpha_1$ fluorescence, whereas the present study investigates the Dy $L\alpha_2$ PFY. Consequently, our results show resonant

Raman scattering from the $L\alpha_1$ in the distant pre-edge region ($E_1 \approx 7760$ eV). The nearer pre-edge feature, a few eV before the main edge, is due to the quadrupole-allowed $2p \rightarrow 4f$ transition, and is present in both studies. These results validate the SWD approach for high-resolution studies of x-ray fluorescence.

**III.C.** Future Directions

Future directions in both instrument development and in applications deserve comment. Starting with the former, if the overall configuration is not near to backscatter, then the aberration inherent to a Johann-style optic subtending a large collection angle will begin to influence spectrometer performance. Different crystals will reflect different bandpasses of radiation through the aperture. At very large deviations from backscatter, crystals near the edge of the optic will not satisfy any Bragg condition with respect to the source and aperture positions.

There are three solutions to this problem. First, one may find a suitable diffracting crystal and plane such that the desired fluorescence energy is relatively close to backscattering. For example, an initial comparison of all known hard x-ray fluorescence energies[23] with allowed reflections from quartz finds that more than 15% of all fluorescence lines could be reflected within 10 degrees of backscatter from suitably oriented crystals.

Second, the optic design can be modified to satisfy the diffraction condition on the Rowland circle, even for large deviations from backscattering. This is the classic approach of DuMond and Kirkpatrick[24] and of Johannson.[25] In the present context, Figure 10 shows a suitable SWD optic.[13] Note the existence of a focal plane, just as for a diced Johann-style analyzer at long working distance.[4] An aperture at this location will select the same energy bandpass from each analyzer crystal, and again leads to unique illumination of the 2D-PSD. Latush, *et al.*,[11] have recently used multiple curved analyzer crystals in this geometry for a focusing spectrometer design. A simple planar optic based on Figure 10 may be useful for diamond anvil cell measurements. A larger collection solid angle will be provided when flat crystals are placed on the surface of revolution formed by rotation of the profile in Fig. 10 about the axis passing through the source and focal points[26] (as shown by the dashed line in the figure). We have begun fabrication of these optics and will report elsewhere on their performance.

Third, the use of flat crystals performing a symmetric Bragg reflection on the Rowland circle is extremely convenient, but may not be optimal. The unique illumination of the 2D-PSD comes at a significant cost. In our prototype instrument, each crystal subtends a square of 9 deg on each edge but only a 5 deg circular cone is selected from each crystal by the aperture so that only 25% of the solid angle of the optic itself is actively participating in spectrometer. It would be interesting to investigate whether the optic's solid angle may be used more efficiently with modestly bent crystals, with asymmetric Bragg reflections from flat crystals, or by other configurations of flat crystals. The logarithmic-spiral geometry (or log-spiral of revolution)[27] or the Wittry geometry[26] are good candidates.

Two other instrumentation-related issues deserve further comment. First, in the present measurements, $\delta\theta_{crystal}$ is much less than $\delta\theta_{source}$ and $\delta\theta_{pixel}$ due to the very high intrinsic energy resolution of the Si 440 reflection for an unstrained crystal. Strained analyzer crystals or analyzer crystals made from materials which are known to have poorer $\delta\theta_{crystal}$ and larger form factors than the traditional, perfect crystals used in bent-crystal optics could give significant improvements over unstrained Si for many measurements. Second, the 2D-PSD used here (Pilatus 100k) was selected because of its high quantum efficiency and the absence of dark current and readout noise. The good time-resolution of this instrument also permits future applications of SWD multicrystal x-ray spectrometers to time-resolved studies. However, for usual, time-integrated measurements of XES for concentrated samples, a high quantum efficiency detector based on a cooled CCD will often be beneficial. When the dark current is sufficiently small, the much larger number of pixels for such detectors ($10^6$ and higher) increases the collection solid angle for given final energy resolution. The choice of energy resolution sets an upper bound on $\delta\theta_{pixel}$, which in turn determines the detector working distance. Therefore, the larger the number of detector pixels, the larger the potential effective collection solid angle. Collection solid angles of 0.5 sr are possible for SWD multicrystal spectrometers based on such detectors when the SWD optic spans a larger portion of a hemisphere than was the case here.

Because of their high efficiency, there is good reason to expect broad application of SWD multicrystal spectrometers. These instruments are convenient for fluorescence detection in usual synchrotron-based spectroscopic techniques, where they should have easy compatibility with many existing endstations for microprobe, diamond anvil-cell, or time-resolved pump/probe measurements. A case of rather extreme parameters is presented by the upcoming commissioning of the Linac Coherent Light Source (LCLS). Based on the experimental conditions and results in the MnO study and the expected availability of a 2D-PSD with $10^6$ pixels at the LCLS x-ray pump-probe endstation, a suitable SWD multicrystal spectrometer should be able to collect high-quality emission spectra with better than 1 eV resolution when a single pulse of the LCLS ($10^{12}$ photons) is incident on relatively concentrated samples.

## V. Conclusions

We present and demonstrate a new approach to the design of dispersing x-ray spectrometers. The optic for our spectrometer is an ensemble of small, flat crystals on a spherical surface at a short working distance to a microfocused source. Each crystal Bragg scatters analyzed radiation onto a 2-dimensional position sensitive detector (2D-PSD), and an aperture ensures that each pixel on the 2D-PSD is uniquely illuminated. Collection solid-angles of 0.1-0.5 sr may be achieved with energy resolution of a few tenths to a few eV in the hard x-ray range. Instruments based on this approach will have many applications at high-brilliance x-ray sources.

**Acknowledgements**

We thank D. Crozier, D. Fritz, A. Macrander, B. Ravel, L. Sorensen, and E. Stern for useful discussions. This research was supported by the Office of Naval Research Grant No. N00014-05-1-0843 and the L. X. Bosack and B. M. Kruger Charitable Foundation. The operation of Sector 20 PNC-CAT/XOR is supported by the U.S. Department of Energy, Basic Energy Science, Office of Science, Contract No. DE-AC02-06CH11357. and grants from the Natural Sciences and Engineering Research Council of Canada. Use of the Advanced Photon Source was supported by the U.S. Department of Energy, Basic Energy Sciences, Office of Science, under Contract No. DE-AC02-06CH11357.

**Figure Captions**

**Figure 1**: (a): A 'primitive' short working distance x-ray spectrometer for a 1 μm spot on the sample. (b) A first step to a compromise design, with lower energy resolution but the potential for much higher collection solid angle through the use of multiple flat crystals. Ray-tracing from the source to the crystals is omitted for clarity. The strong overlap of backscattering cones prevents unique illumination of the 2-dimensional position sensitive detector (2D-PSD) by the two crystals. (c) The use of an aperture allows unique illumination of the 2D-PSD by multiple flat analyzer crystals. Ray-tracing from the source to crystals 2 through 5 is omitted for clarity. See the text for discussion.

**Figure 2**: A section view from above of a prototype short working distance multicrystal x-ray spectrometer. Labeled components are: *a*- 2.54 cm diameter optic which supports 22 flat Si 110 crystals; *b*- cylindrical spacer; *c*- beam entrance; *d*- beam exit; *e*- aperture plate; *f*- retaining ring. The outer support structure consists of standard lens tube components.

**Figure 3**: (a) A photograph of a 22-element short working distance x-ray optic. The inner-diameter of the support tube is 25.5 mm. (b) A 10 second exposure at an incident energy of 7800 eV, showing the strong Dy fluorescence occurring near backscatter for the Si (440) analyzer crystals. The rectangular protrusion in the reflection from each crystal is the shadow of the sample holder.

**Figure 4**: (*a*)-(*d*): A small portion of the detector output during energy calibration scans for the MnO experiment; (*e*): A contour plot of the resulting calibration. The lowest (rightmost) contour is 6470.0 eV and successive contours are spaced by 2.0 eV. (*f*): A measurement of Mn Kβ fluorescence from MnO. The black ovals show the boundary of the shadow of the spectrometer's aperture. The indicated energies in a panel give the incident photon energy $E_1$. The exposure time is 120 sec for the calibration scans and 40 sec for the fluorescence measurement.

**Figure 5**: (*a*)-(*d*): A small portion of the detector output during energy calibration scans for the Dy experiment; (*e*): A contour plot of the resulting calibration. The smallest semicircular contour is 6458.0 eV and successive contours are spaced by 1.0 eV, increasing as one moves to contours with larger radii of curvature. (*f*): A measurement of Dy $K\alpha_2$ fluorescence from Dy foil. The gray ovals show the boundary of the shadow of the spectrometer's aperture. The indicated energies in a panel give the incident photon energy $E_1$. The exposure time is 10 sec for each data panel.

**Figure 6**: The Mn Kβ emission spectrum for MnO measured at an incident energy $E_1 = 6558.0$ eV with 1.35 eV resolution. The total measurement time is 40 sec, and only 30% of the analyzer crystals are used in the measurement. The incident flux is 5 $10^{11}$/sec.

**Figure 7**: Contour plot of the Mn Kβ RIXS from MnO. The horizontal dashed lines indicate the position of the $K\beta_{1,3}$ (upper) and $K\beta'$ (lower) peaks in the emission

spectrum. The diagonal dashed line shows a path of constant energy transfer going through the most intense feature in the $K\beta_{1,3}$ XANES.

**Figure 8**: $K\beta'$ and $K\beta_{1,3}$ partial fluorescence yield XANES measurements for MnO. The inset shows an enlargement of the energy region where $1s \rightarrow 3d$ transitions contribute to the $K\beta_{1,3}$-based spectrum.

**Figure 9**: Partial fluorescence yield XANES measurements of Dy foil near the $L_3$ absorption edge at emission energies corresponding to the $L\alpha_2$ fluorescence line. The spectra have been vertically displaced for clarity of presentation. The center of the detection ($E_2$) bandpass shifts 0.5 eV between successive curves. The dashed lines are curves of constant energy loss for the resonant Raman scattering for the $L\alpha_1$ fluorescence and the $2p \rightarrow 4f$ quadrupole transition.

**Figure 10**: Ray-tracing for a short working distance multicrystal spectrometer patterned on the criterion of DuMond and Kirkpatrick[24] and Johannson[25] for correct operation on the Rowland circle. As constrained by the aperture, each Bragg-scattered cone subtends 4.5 degrees of arc. The centers of successive crystals are spaced by 9 degrees. Note the existence of a true focal plane, as in Huotari, *et al.*,[4] for a long working distance, diced Johann optic. The nonfocusing geometry shown in the figure results in weak tolerances on assembly and operation.

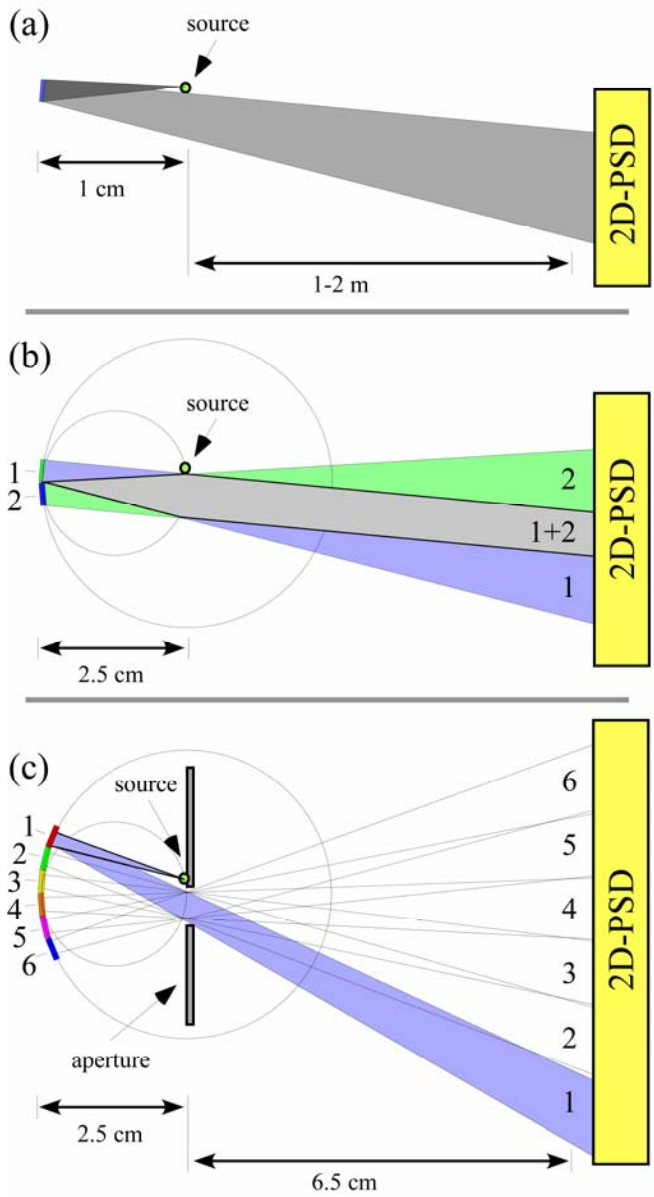

**Figure 1:** B. Dickinson, *et al*., submitted, Rev. Sci. Instrum. 2008.

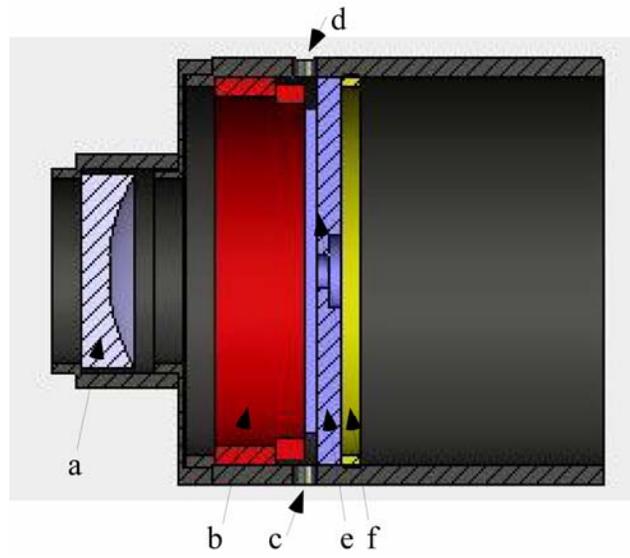

**Figure 2:** B. Dickinson, *et al*., submitted, Rev. Sci. Instrum. 2008.

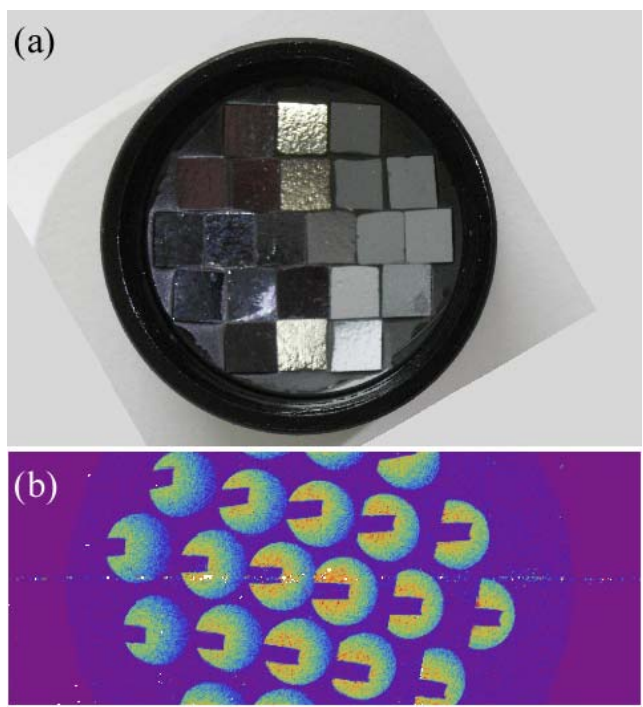

**Figure 3:** B. Dickinson, *et al*., submitted, Rev. Sci. Instrum. 2008.

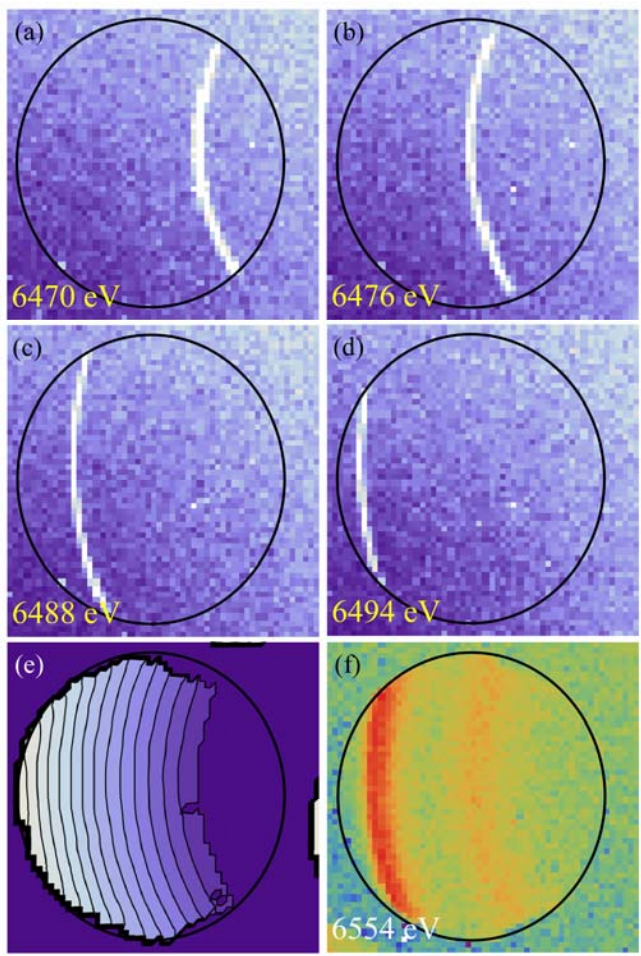

**Figure 4:** B. Dickinson, *et al*., submitted, Rev. Sci. Instrum. 2008.

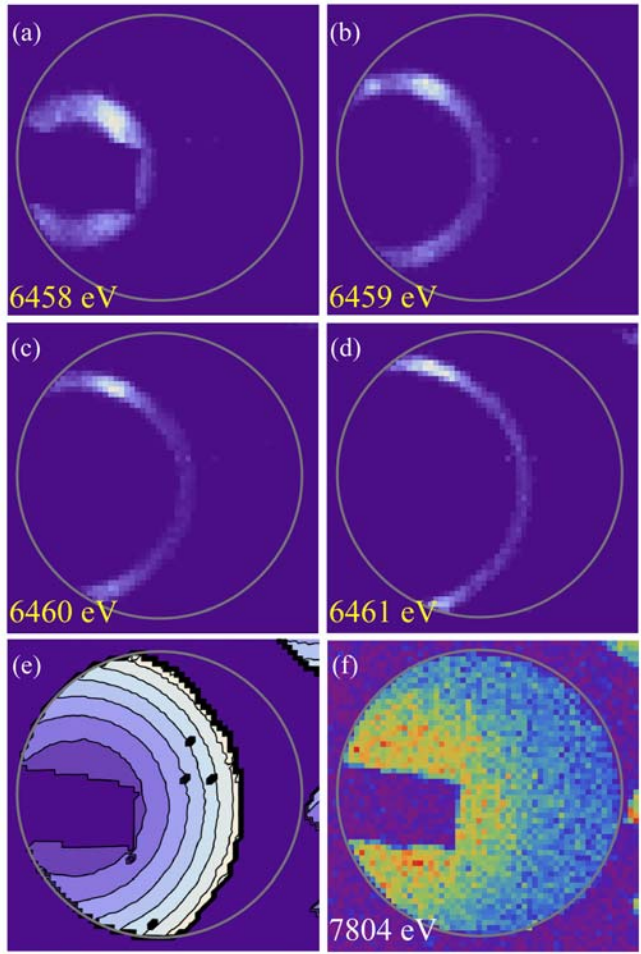

**Figure 5:** B. Dickinson, *et al*., submitted, Rev. Sci. Instrum. 2008.

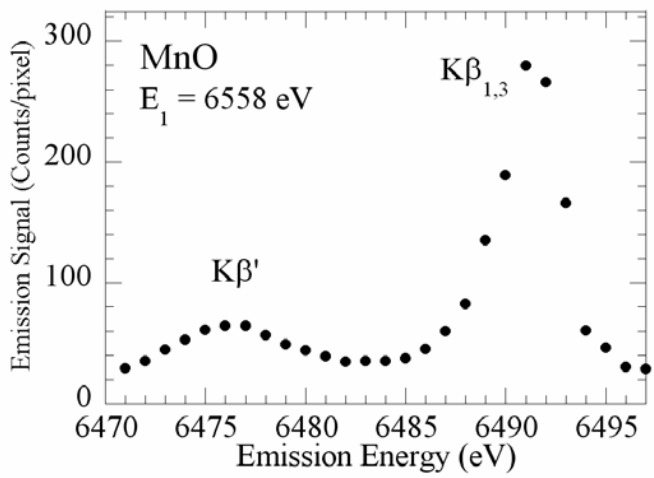

**Figure 6:** B. Dickinson, *et al*., submitted, Rev. Sci. Instrum. 2008.

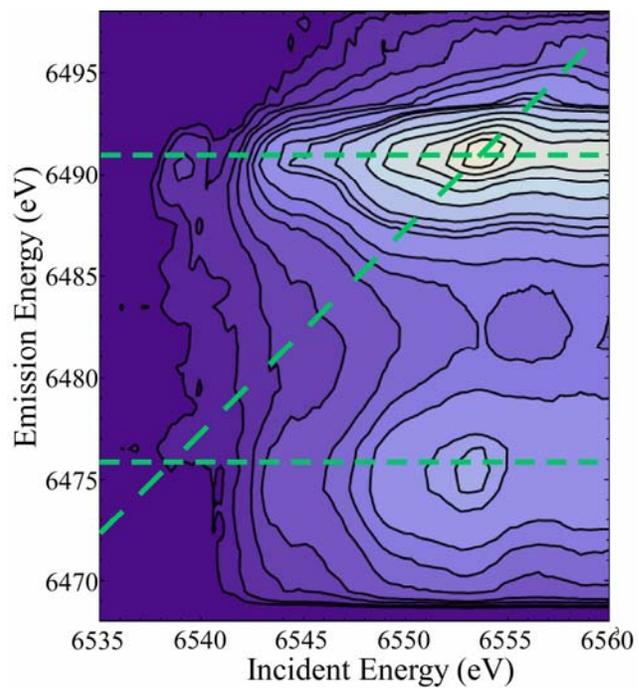

**Figure 7:** B. Dickinson, *et al*., submitted, Rev. Sci. Instrum. 2008.

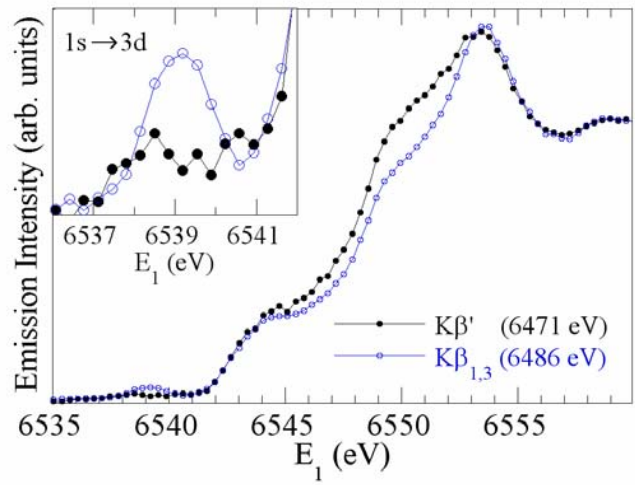

**Figure 8:** B. Dickinson, *et al*., submitted, Rev. Sci. Instrum. 2008.

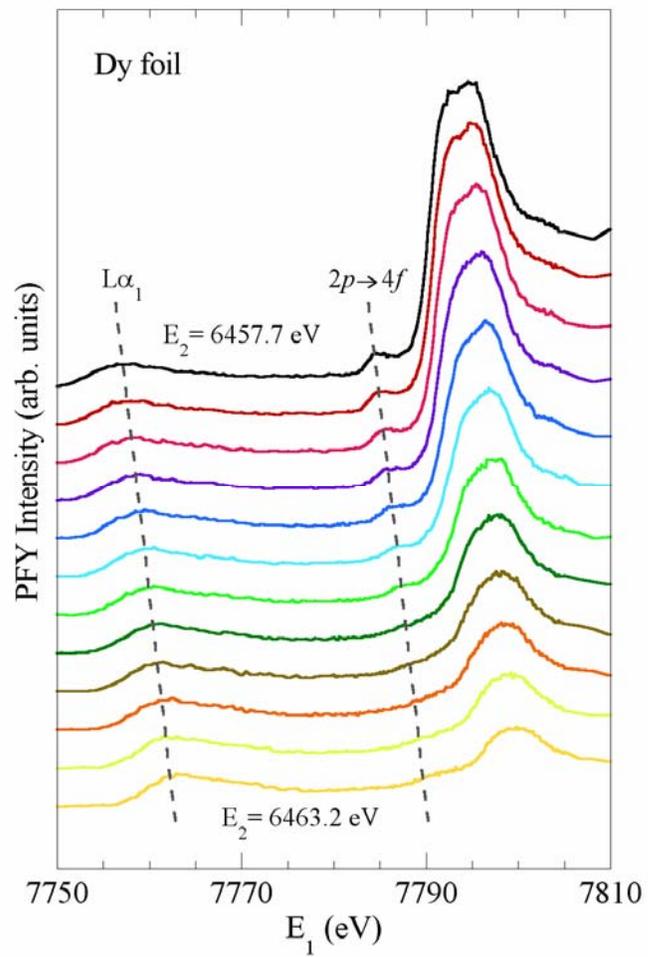

**Figure 9:** B. Dickinson, *et al*., submitted, Rev. Sci. Instrum. 2008.

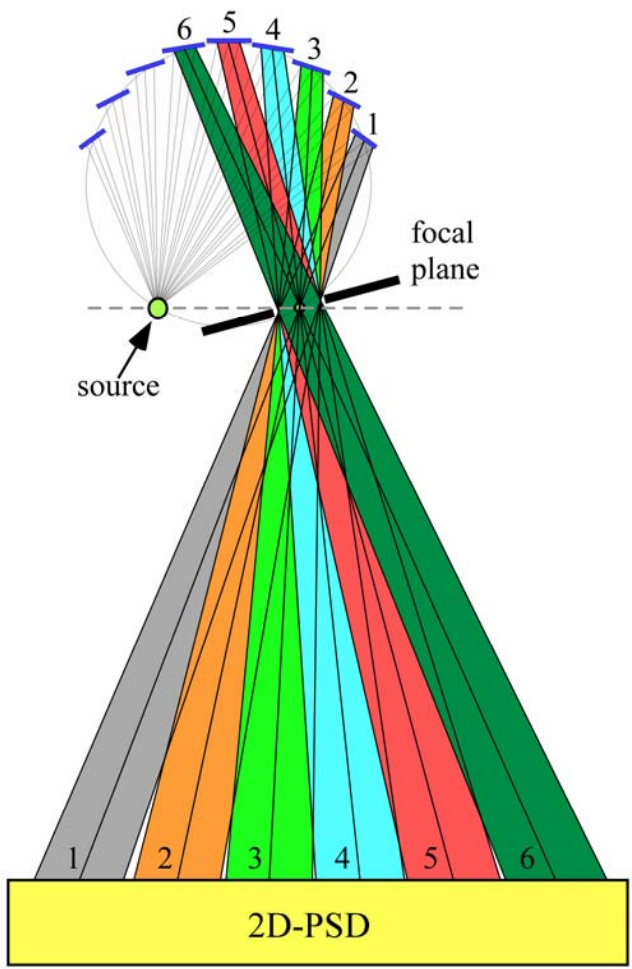

**Figure 10:** B. Dickinson, *et al*., submitted, Rev. Sci. Instrum. 2008.